\documentclass[]{spie}  

\usepackage{amsmath,amsfonts,amssymb}
\usepackage{graphicx}
\usepackage[table,xcdraw]{xcolor}
\usepackage{siunitx}
\usepackage[colorlinks=true, allcolors=blue]{hyperref}

\DeclareMathOperator*{\argmin}{arg\,min}

\title{The Space Coronagraph Optical Bench (SCoOB): 11. Modeling and correction of chromatic aberrations}

\author[a]{Kyle Van Gorkom}
\author[a]{Ramya M.~Anche}
\author[a]{Saraswathi Kalyani Subramanian}
\author[a,b]{William Melby}
\author[a]{Kian Milani}
\author[a,b]{Emory Jenkins}
\author[a]{Irina Stefan}
\author[a]{Adam Schilperoort}
\author[a]{Patrick Ingraham}
\author[c]{Jaren N. Ashcraft}
\author[a,b]{Kevin Derby}
\author[a,b]{Daewook Kim}
\author[b]{Heejoo Choi}
\author[a]{Olivier Durney}
\author[a]{Ewan S. Douglas}

\affil[a]{Steward Observatory, University of Arizona, Tucson, AZ}
\affil[b]{James C. Wyant College of Optical Sciences, University of Arizona Tucson, AZ}
\affil[c]{Department of Physics, University of California, Santa Barbara, CA}

\authorinfo{Further author information: (Send correspondence to K.V.G.)\\K.V.G.: E-mail: kvangorkom@arizona.edu}

\begin{document} 
\maketitle

\begin{abstract}

The space coronagraph optical bench (SCoOB) is a high contrast imaging testbed designed to demonstrate starlight suppression techniques at visible wavelengths in a space-like vacuum environment. Since the previous proceedings, the testbed has undergone major component upgrades, including a 100\%-yield BMC Kilo-C deformable mirror, a black silicon pupil stop, and circular polarizers between wedged substrates. To assess the performance limits imposed by lateral chromatic aberration induced by refractive optical elements, we develop analytic and numerical tools to predict the broadband contrast. Additionally, we present the broadband contrast performance of the testbed using implicit electric field conjugation with multiple sensing bands for broadband control.

\end{abstract}


\section{INTRODUCTION}

\begin{figure}
    \centering
    \includegraphics[width=0.5\linewidth]{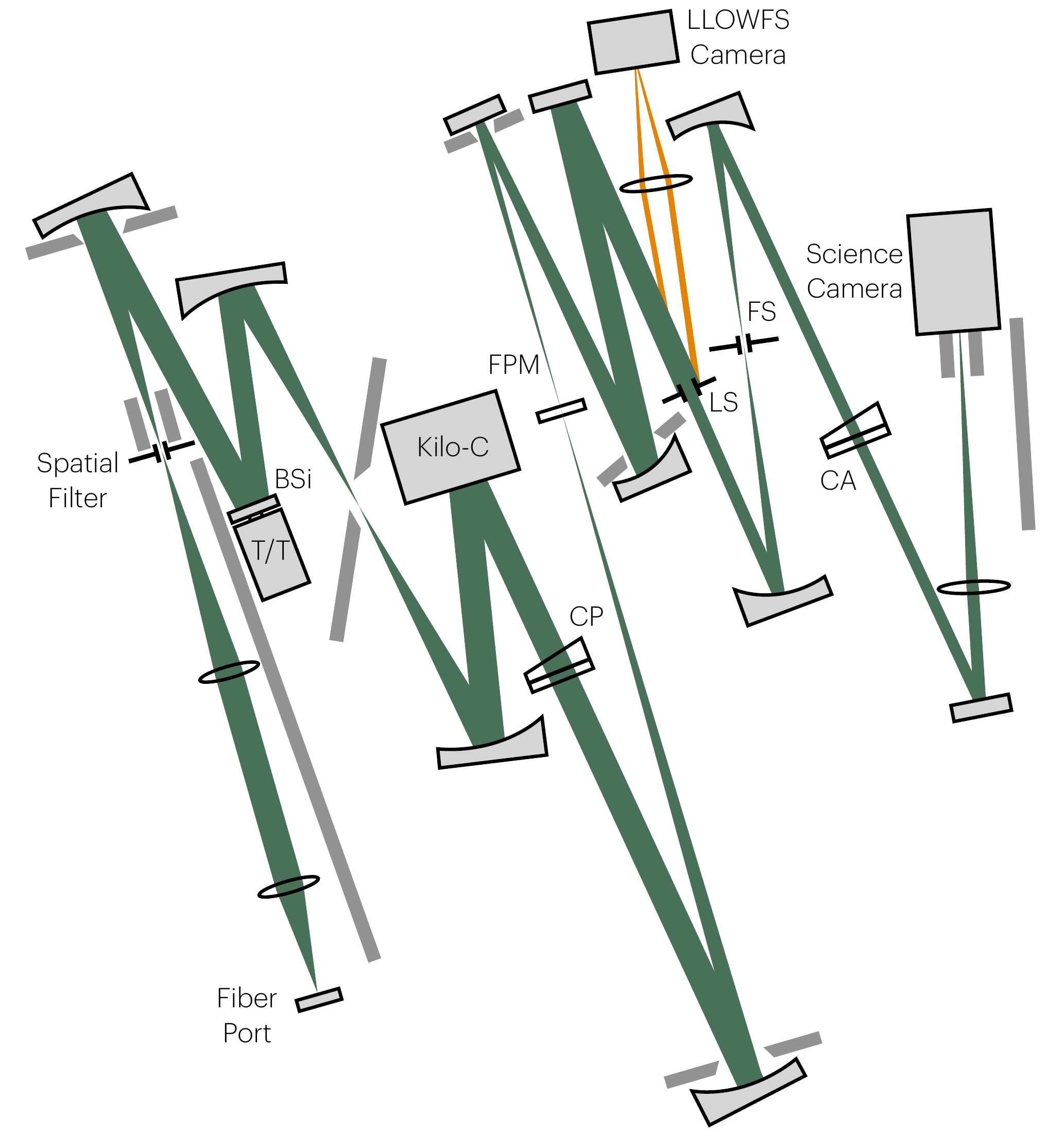}
    \caption{Current layout of SCoOB. Notable recent changes include an upgraded black silicon (BSi) pupil stop, a 100\% yield Al-coated Kilo-C DM, reflective Lyot stop (LS), and circular polarizers (CA, CP) between wedged substrates. A handful of baffles (shown as gray blocks) have been added in an attempt to mitigate stray light. The focal plane mask (FPM) is a charge-6 VVC in an \({\sim}\)f/51 beam with a 45\si{\micro \meter}-diameter opaque dot. A field stop (FS) is included in the layout but not present for the experiments described here.}
    \label{fig:testbed_sketch}
\end{figure}

The Space Coronagraph Optical Bench (SCoOB)\cite{maier_2020,ashcraft_2022} is a thermal vacuum (TVAC) high contrast imaging testbed that combines a charge-6 vector vortex coronagraph (VVC) with a 34x34 Kilo-C microelectromechanical systems (MEMS) deformable mirror from Boston Micromachines Corp.~(BMC). SCoOB previously demonstrated a contrast performance of $2.2\times10^{-9}$ in a $\ll 1 \%$ bandwidth (BW), $4\times10^{-9}$ in a 2\% BW, and $2.5\times10^{-8}$ in a 15\% BW\cite{vangorkom_2024}. Previous studies have discussed end-to-end diffraction modeling of the testbed\cite{anche_2025}, simulation and measurement of polarization aberrations\cite{anche2024space,ashcraft2024space}, demonstration of Lyot-Low Order Wavefront Sensing (LLOWFS)\cite{milani2025space}, and the design, fabrication, and initial operation of a self-coherent camera\cite{derby2025space}. Driven by previous analysis of our contrast limits\cite{anche_2025}, we recently upgraded to a 100\%-yield Al-coated Kilo-C, a black silicon (BSi) pupil stop, circular polarizers between wedged substrates (for ghost mitigation), and added Acktar-coated baffles throughout to mitigate stray light. The SCoOB layout and a description of the major components of the testbed are given in Figure \ref{fig:testbed_sketch}.

In these proceedings, we focus in particular on the chromatic aberrations induced by transmissive optics and develop analytic and numerical models to predict the broadband contrast floor set by dispersion-driven effects. We report the latest broadband performance of SCoOB with wedged substrates in the beam path with sensing and correction in multiple sensing bands. Finally, we give an update on the status of our end-to-end modeling tools intended to inform the limits of our testbed performance.

\section{Lateral Chromatic Aberration In High Contrast Imaging Systems}\label{sec:lca}

Transmissive optical elements are notorious sources of stray/scattered light and chromatic wavefront error (WFE). While coronagraph instruments attempt to minimize the use of such optics to avoid the contrast limits imposed by these effects, these optics can't be entirely avoided---common examples include dichroics, filters, polarization optics, and some coronagraph masks. When refractive optics are downstream of the coronagraph masks or a field stop, the impact to the contrast tends to be negligible; when placed earlier in the optical system, however, these optics may limit the achievable contrast.

For VVCs, polarization filtering is essential to reduce the polarization leakage term to acceptable levels \cite{mawet_2010}. This is typically done via linear polarizers and quarter wave plates sandwiched between transmissive substrates. Due to imperfect antireflection (AR) coatings on these substrates, a consequence of this is a ``ghost'' at the \({\sim}10^{-6}\) contrast level (two 0.2\% reflections). In a testbed, this ghost can be rejected by placing the input circular polarizing optics upstream of a spatial filter, but this solution is not available to a flight-like instrument.

One mitigation strategy in a flight-like configuration is to add a small wedge angle to the substrates to moves the ghost PSF off-axis; the consequence of this however, is the introduction of lateral chromatic aberration. In the sections that follow, we develop analytic estimates of the broadband contrast floor imposed by dispersive wedges. In addition, because dichroics are of general interest to the community, we analyze contrast floor imposed by plane-parallel plates (PPPs) at oblique angles. Finally, we report our experimental contrast on SCoOB in the presence of wedged substrates upstream of the VVC.

\subsection{Wedged Substrates}

Adding small wedges to transmissive optics is a common technique to move back-reflected ``ghosts'' off-axis. With a wedge angle $\alpha$ on a substrate with a refractive index \( n \),  the angular deviation of the primary ray through a thin prism is given by \( \delta = \left[ n - 1 \right] \alpha \). The secondary ghost ray (which sees two reflections at the substrate interfaces) is deviated by an additional \( \Delta \delta = 2 n \alpha \) with respect to the primary ray. For a beam size $D$ at the substrate location and a dark hole with extent  \( s_\mathrm{OWA} \lambda/D \), the wedge angle that will move the ghost to the edge of the DH is
\begin{equation}
	\alpha_\mathrm{min}	= \frac{s_\mathrm{OWA}}{2 n} \lambda /D.
\end{equation}

In practice, some larger wedge angle (e.g., \( 2 \alpha_\mathrm{min} \)) would be chosen to ensure that the ghost PSF is well outside the outer working angle.

\begin{figure}
    \centering
    \includegraphics[width=0.7\linewidth]{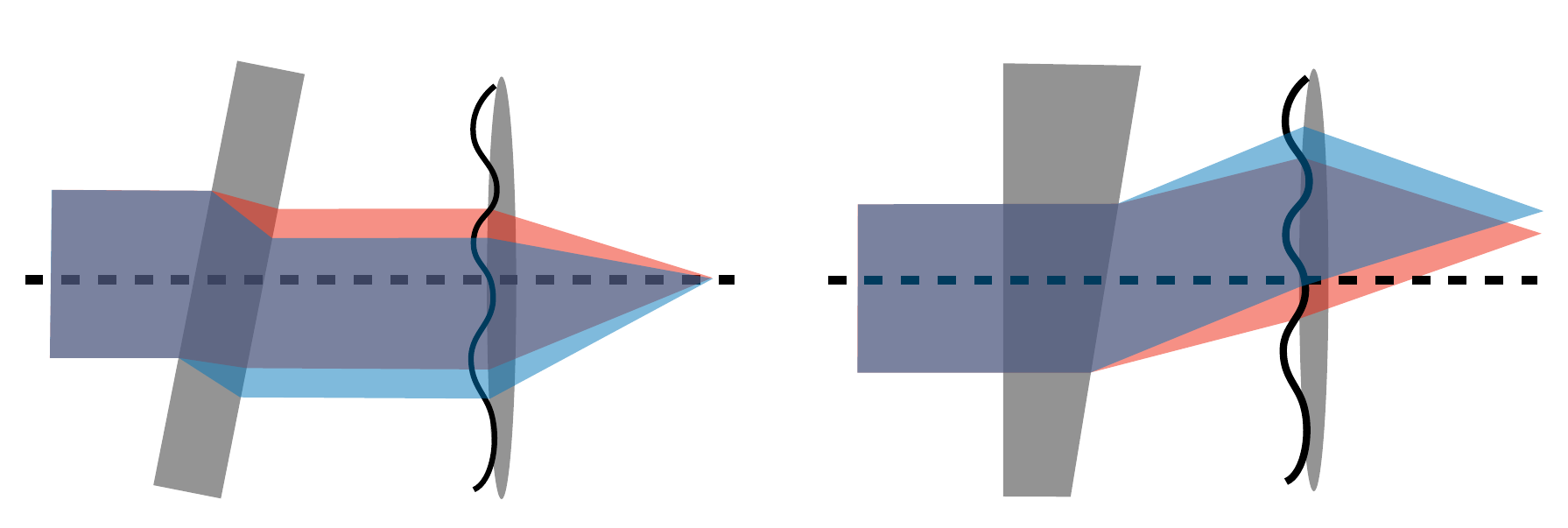}
    \caption{Left: illustration of chromatic beam shear arising from a tilted plane parallel plate upstream of an optic with surface figure error. Right: chromatic beam shear and chromatic displacement in the focal plane arising from a wedged substrate.}
    \label{fig:lca_illustration}
\end{figure}

  One consequence of a putting a wedge on a substrate, however, is that it acts as a prism due to the chromaticity of the refractive index. The angular dispersion at wavelength $\lambda$ with respect to a reference wavelength $\lambda_0$ is given by
  
\begin{equation}
    \Delta \delta(\lambda, \lambda_0) = \left[ n(\lambda) - n(\lambda_0) \right] \alpha \approx  (\lambda - \lambda_0) \frac{\delta n}{\delta \lambda_0} \alpha,
\end{equation}

where $\delta n/\delta \lambda_0$ is the chromatic dispersion of the substrate at the reference wavelength. This dispersion results in two related but distinct effects.

The first is a lateral chromatic displacement of the PSF at the focal plane mask, which results in an uncorrectable tip/tilt error that varies with wavelength. Over a bandwidth \( B = \Delta \lambda / \lambda_0 \), a wedged substrate in a plane with a beam footprint of diameter $D$ displaces the PSF by a total range of
\begin{equation}
 s_{\lambda_0/D} (B) = \alpha B D  \frac{\delta n}{\delta \lambda_0}
\end{equation}
in units of \( \lambda_0/D \). For example, a 30 arcminute wedge on an N-BK7 substrate with an approximate chromatic dispersion of \( \delta n/ \delta \lambda_0 \approx -0.029\  \si{\micro \meter}^{-1} \) at 630 nm in a 10 mm beam gives a focal-plane displacement of \( \pm 0.13\  \lambda_0/D \) in a 10\% bandwidth.  The impact on contrast is heavily dependent on the coronagraph architecture: coronagraphs that rely primarily on pupil-plane masks are expected to be largely insensitive to this, while focal-plane coronagraph masks should be significantly more susceptible.

\begin{figure}
    \centering
    \includegraphics[width=0.5\linewidth]{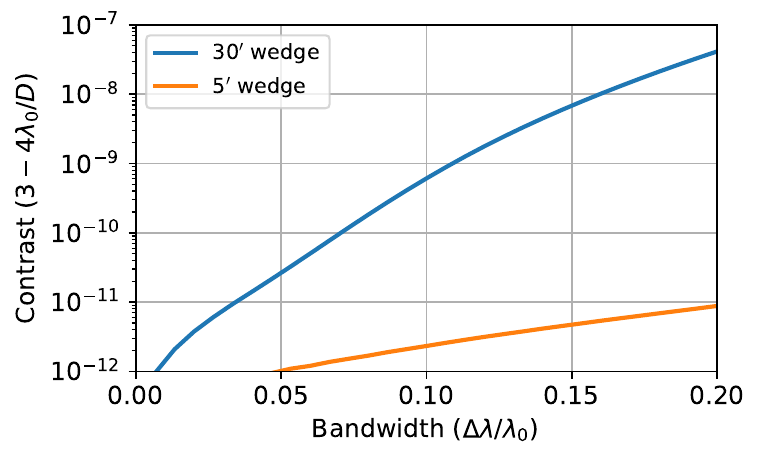}
    \caption{Contrast at \(3-4\lambda_0/D\) vs bandwidth due to lateral chromatic aberration at a charge-6 VVC for an optic with a 5 and 30 arcminute wedge. Values plotted here capture the effect of the chromatic displacement of the PSF at the FPM alone and do not include chromatic beam shear.}
    \label{fig:chromatic_disp_vvc6}
\end{figure}

To quantify this effect for SCoOB, we numerically simulated the contrast at $3\text{-}4\ \lambda_0/D$ for a charge-6 VVC in bandwidths up to 20\% centered at 630 nm. These results are plotted in Figure \ref{fig:chromatic_disp_vvc6}. The contrast floor set by a 5 arcminute wedge remains below \(10^{-11}\) up to 20\% BW, while a 30 arcminute wedge exceeds \( 10^{-10} \) contrast for bandwidths \( {>}10\% \).

The second consequence of the angular dispersion is a chromatic beam shear on surfaces downstream of the wedge, which results in uncorrectable, chromatic mid- to high-spatial-frequency errors in the wavefront. The impact on contrast can be evaluated analogously to beamwalk\cite{mendillo_2017, vangorkom_2025}. Assuming a perfect DM correction for a sinusoidal surface error at the central wavelength, the chromatic aberration at other wavelengths can be treated as a phase shift on that error. Let an optic a distance \(z\) away from the wedged substrate have a phase aberration given by
\begin{equation}
	\phi = \beta_k \frac{4 \pi}{\lambda}	\sin \left( 2 \pi \frac{\vec{k} \cdot \vec{r}}{D} \right),
\end{equation}
where \(\vec{k}\) is the spatial frequency of the error in cycles/aperture, and \(\vec{r}\) is the lateral coordinate vector. The phase shift on this sinusoidal error seen at an off-central wavelength \(\lambda\) is
\begin{equation}
	\Delta \phi = 2\pi \frac{\delta n}{\delta \lambda_0} \left( \lambda - \lambda_0 \right) \alpha z \frac{k}{D} \cos{\gamma},
\end{equation}
where \(k\) is the radial spatial frequency and \(\gamma\) is the angle between the clocking of the wedge \(\alpha\) and the spatial frequency \(\vec{k}\) of the sinusoidal error. At this wavelength, the normalized focal-plane intensity is
\begin{equation}
    I(k, \gamma; \lambda) = \beta_k^2 \frac{16 \pi^2} {\lambda^2} \sin^2\left( \pi \frac{\delta n}{\delta \lambda_0} \left[ \lambda - \lambda_0 \right] \alpha z \frac{k}{D} \cos \gamma \right) 
\end{equation}

Then the mean contrast over a bandwidth \(B\) assuming a flat stellar spectrum is found by integrating over the bandpass:
\begin{equation}
\begin{aligned}
    C(k, \gamma; B)  &= \beta_k^2 \frac{16 \pi^2}{\Delta \lambda} \int_{\lambda_0 - \Delta \lambda/2}^{\lambda_0 + \Delta \lambda/2} \frac{\sin^2\left( \pi \frac{\delta n}{\delta \lambda_0} \left[ \lambda - \lambda_0 \right]\alpha z \frac{k}{D} \cos \gamma \right) } {\lambda^2}  d \lambda \\
        & \approx \beta_k^2 \frac{4 \pi^4}{3} \left(\frac{\delta n}{\delta \lambda_0}\right)^2 \left(\frac{k}{D}\right)^{2} \alpha^2 z^2 \cos^2 \gamma B^2,
\end{aligned}
\end{equation}
where the integral is evaluated by retaining the first-order terms in the Taylor series for the integrand, and the chromatic dispersion \(\delta n/ \delta \lambda_0\) is taken to be a constant over the bandwidth of interest. Note that the form of this integral is functionally identical to the expression for the broadband contrast floor due to the Talbot effect in the first-order phase approximation\cite{Shaklan06, mazoyer_pueyo, vangorkom_2025}.

With a simple 2D isotropic power-law power spectral density (PSD) of the form \(\mathrm{PSD}(k) = \beta (k/D)^{-c} \) to represent the surface errors on the downstream optic, the contrast floor at a radial separation $k \lambda/D$ and angle \( \gamma \) can be expressed as
\begin{equation}\label{eqn:c_wedge}
    C_{\mathrm{wedge}}(k, \gamma; B) = \beta \frac{8 \pi^4}{3} \left(\frac{dn}{d\lambda}\right)^2 \alpha^2 z^2 B^2 \cos^2 \gamma \frac{k^{2-c}}{D^{4-c}},
\end{equation}
where the variance \(\beta_k^2\) at \(\vec{\nu} = \vec{k}/D\) cycles/aperture has been replaced by \( \mathrm{PSD}(\nu) \nu d\nu d\gamma  = \dfrac{\mathrm{PSD}(k/D)}{D^2} k dk d\gamma\) to account for the 2D PSD, the contrast quantity is normalized by the differential area element \(  C_{\mathrm{wedge}}(k, \gamma; B) \equiv \dfrac{C(k, \gamma; B)} {k dk d\gamma} \), and an extra factor of two is introduced to account for the combined contrast contribution from the power at both the \( +\vec{k} \) and \( -\vec{k} \) spatial frequencies. Note that this simplified analysis ignores the contribution from spatial frequency cross-terms (frequency folding).

An example of this is plotted in Figure \ref{fig:c_wedge_ppp} for a 5 and 30 arcminute wedge in a beam with a 10 mm diameter and a surface error (aligned with the wedge clocking) a distance  \(z = 300\) mm downstream of the wedged substrate in a 10\% bandwidth. The PSD on this surface is assumed to follow an ABC PSD of the form
\begin{equation}
	\mathrm{PSD}(k) = \frac{a}{1 + \left(\frac{k/D} {b} \right)^c}	
\end{equation}
with \( a = 9.3\times10^{-5} \) nm\(^2\) m\(^2\) (4.7 nm RMS surface), \( b = 154\  \mathrm{m}^{-1} \) (\({\sim}2.5\) cycles/aperture), and \( c=2.7\), values that were found to be a good fit to measured surface figure PSDs for SCoOB off-axis parabolas (OAPs)\cite{vangorkom_2025}. The contrast floor with this PSD reduces to the simple power-law expression in Equation \ref{eqn:c_wedge} for $k \gg 2.5$.

The 5 arcminute wedge is again found to have a fairly negligible contribution to the contrast floor in a 10\% bandwidth, but a 30 arcminute wedge exceeds \(10^{-10}\) contrast at separations \( {<}15\  \lambda_0/D \).

To test the impact of wedged substrates on SCoOB, we worked with Bolder Vision Optics (BVO) to fabricate integrated circular polarizers sandwiched between two wedged substrates, with the wedge angle on the input and output substrates clocked at \(90^{\circ}\) with respect to each other. The initial prototypes tested here have 30 arcminute wedges such that the total effective wedge was \( {\sim} 42^{\circ} \), but future units will have wedge angles closer to 5 arcminutes. Testbed results with these substrates are discussed in Section \ref{sec:iefc}.

\subsection{Plane-Parallel Plates}

To estimate the impact of lateral chromatic aberration that arises from transmissive optics placed at non-zero angles of incidence (AOIs), we analyze the case of a plane-parallel plate (PPP). Examples of tilted PPPs in coronagraph instruments include dichroics and any other transmissive elements intentionally tilted to redirect the reflection from the front interface to a beam dump to reduce stray light. We note, however, that this analysis does not include impact of the polarization effects or chromatic wavefront error (WFE) that arises from the thin-layer stack of the dichroic. An analysis of a dichroic design for a high contrast imaging application is presented in Rosenbluth et al., these proceedings\cite{rosenbluth_2026}, and measurements of a commercial dichroic in SCoOB are presented in Cooper et al., these proceedings\cite{cooper_2026}.

A plane parallel plate in a collimated beam at an angle \(\theta\) with respect to the optical axis creates a chromatic beam shear without angular deviation (see Figure \ref{fig:lca_illustration}). The displacement of the shear is given by
\begin{equation}
    d \approx t \theta \left( \frac{n(\lambda) - 1} {n(\lambda)}\right),
\end{equation}
where \(t\) is the thickness of the substrate. The linearized chromatic change in the displacement with respect to a given central wavelength is
\begin{equation}
    \Delta d = \frac{t \theta}{n(\lambda_0)^2} \frac{\delta n}{\delta \lambda_0} \left( \lambda - \lambda_0 \right) \cos \gamma,
\end{equation}
where \( \gamma \) is the angle between the phase error and the beam deviation.

\begin{figure}
    \centering
    \includegraphics[width=0.5\linewidth]{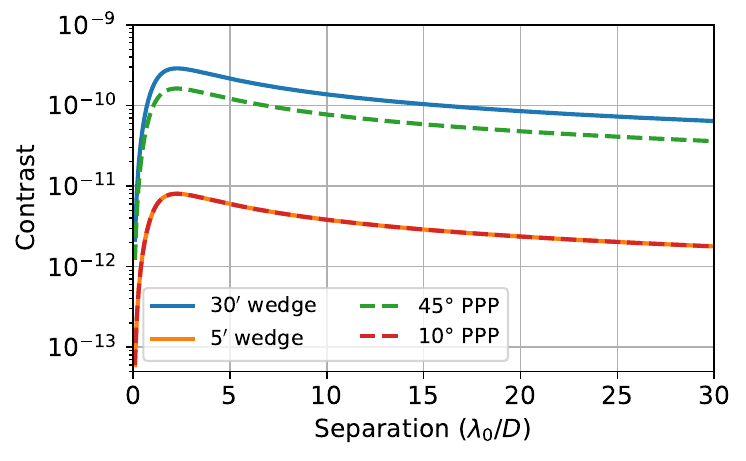}
    \caption{Contrast floor in a 10\% BW set by chromatic beam shear arising from wedged substrates (solid lines) and plane parallel plates (dashed lines) angled with respect to the oncoming beam. Figure errors follow an ABC PSD with parameters given in the text. Note that the contrast values plotted here are roughly representative of performance expected on SCoOB, but Equations \ref{eqn:c_wedge} and \ref{eqn:c_ppp} show that the contrast floor is highly sensitive to instrument design parameters.}
    \label{fig:c_wedge_ppp}
\end{figure}

Following an analysis similar to that above, the chromatic phase shift on a downstream optic (now independent of the distance to that optic) will be 
\begin{equation}
	\Delta \phi = 2 \pi \frac{t \theta}{n(\lambda_0)^2} \frac{\delta n}{\delta \lambda} \frac{k}{D} \left( \lambda - \lambda_0 \right) \cos \gamma
\end{equation}
This has an identical dependence on wavelength as the expression for the broadband contrast floor due to a wedged optic, so we can just write down the corresponding broadband contrast floor:
\begin{equation}\label{eqn:c_ppp}
    C_{\mathrm{PPP}}(B, k, \theta) = \beta \frac{8 \pi^4}{3} \left(\frac{\delta n}{\delta \lambda_0}\right)^2 \left( \frac{t \theta}{n(\lambda_0)^2} \right)^2 B^2  \cos^2\gamma \frac{k^{2-c}}{D^{4-c}},
\end{equation}
where we've again assumed an isotropic power-law PSD on a downstream surface. Note that this expression is not strictly valid for the larger AOIs typical for, e.g., a dichroic, but since it produces a slight overestimate of the contrast floor compared to the exact expression, we use it regardless.

An example of this effect is plotted in Figure \ref{fig:c_wedge_ppp} for 6mm-thick PPPs placed at \(10^{\circ}\) and \(45^{\circ}\) AOIs in a beam with a 10mm diameter, again with an ABC PSD figure error derived from SCoOB measurements on the downstream optic. The \(10^{\circ}\) remains well below \( 10^{-10} \) contrast, while the \(45^{\circ}\) case limits the contrast to \( {\sim}10^{-10} \).

\section{MULTI-BAND IEFC}\label{sec:iefc}

Previous work shows that performing electric field conjugation (EFC) with pairwise probing at multiple wavelengths can yield improved broadband contrast\cite{giveon_broadband_2007,ruane_broadband_2022}. Implicit electric field conjugation (iEFC)\cite{haffert_implicit_2023} can be directly run on broadband coronagraphic images, but previous experience on the SCoOB testbed has shown that the broadband contrast is degraded compared to performing iEFC on narrowband images. With a minor extension to iEFC, we find the multi-band iEFC control law found by solving
\begin{equation}
    \hat{u} = \argmin_{u} \left\{ \sum_i |\phi_{\lambda_i} + G_{\lambda_i} u|^2 + \mu |u|^2  \right\},
\end{equation}
where $u$ is the vector of DM commands, $\mu$ is a regularization parameter, $\phi_{\lambda_k}$ is the vector of pairwise probed images at the $\lambda_i$ wavelength, and $G_{\lambda_i}$ is the iEFC response matrix at the $\lambda_i$ wavelength.

\begin{figure}
    \centering
    \includegraphics[width=0.5\linewidth]{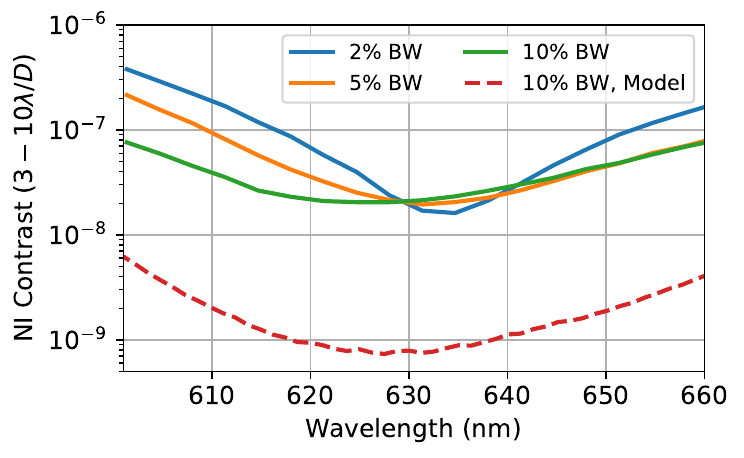}
    \caption{Measured contrast vs wavelength on SCoOB with multi-band iEFC in 2\% (1-band), 5\% (2-band), and 10\% (3-band) bandwidths, compared to the end-to-end model prediction for the broadband contrast floor.}
    \label{fig:contrast_scoob}
\end{figure}

We compare the broadband performance of this algorithm on SCoOB over 3 scenarios: (a) \(i=3\) 2\% sub-bands distributed over a total 10\% bandwidth centered at 630 nm, (b) \(i=2\) 2\% sub-bands distributed over a 5\% bandwidth, and (c) a \(i=1\) 2\% band. After digging with iEFC in each scenario, the broadband performance is measured in 20 (partially-overlapping) 2\% bands spanning a 10\% bandwidth.  These experimental results are compared to performance predicted by an end-to-end model simulating 3 sub-band correction (i.e., scenario a). See Figure \ref{fig:contrast_scoob}. With wedged substrates in the beam path, multi-band iEFC achieves \({\sim}4\times10^{-8}\) contrast in a 10\% BW and \({\sim}3\times10^{-8}\) in a 5\% BW. Prior to the install of wedged substrates, SCoOB achieved \({\sim}2\times10^{-8}\) contrast in a 10\% bandwidth and \({\sim}8\times10^{-9}\) in a 5\% BW\cite{vangorkom_2024}. The simulated 3-band 10\% contrast achieves \({\sim}3\times10^{-9}\) contrast, so there is a significant discrepancy in the predicted performance compared to the actual testbed performance. This suggests there is either an additional physical effect that limits the contrast beyond the terms analyzed in Section \ref{sec:lca}, or else an underestimation of the figure error on the post-CP OAP (which couples into the chromatic beam shear to set broadband the contrast floor). The latter can be investigated in a future experiment which moves the CP closer to this OAP to reduce the amplitude of the shear. Per Equation \ref{eqn:c_wedge}, the broadband contrast scales as \(z^2\), so even a modest reduction in this distance should produce a measurable improvement in the contrast.

The state of the testbed during the experiments described here is summarized in the \texttt{scoob\_state} github respository at \href{https://github.com/uasal/scoob_state/commit/de8263fc5c8b9a0f29dc80a5066ffe5676309ae8}{scoob\_state@de8263f}.

\section{DIFFRACTION MODELING}

To support testbed development and exploration of contrast limits, we developed a GPU-accelerated \texttt{POPPY} (Physical Optics Propagation in PYthon)\cite{poppy} model that can be dynamically generated from a parametrized .toml file. The model integrates:

\begin{itemize}
    \item Fresnel propagation between every surface, with statistically-generated reflectivity and surface errors
    \item Vector diffraction, polarization aberrations, and polarization optical elements
    \item Lateral chromatic aberration
    \item DM quantization and noise
    \item Dynamic WFE, including jitter, beamwalk, and 6-DOF rigid body motion of individual optics in the paraxial regime
\end{itemize}

\begin{figure}
    \centering
    \includegraphics[width=0.6\linewidth]{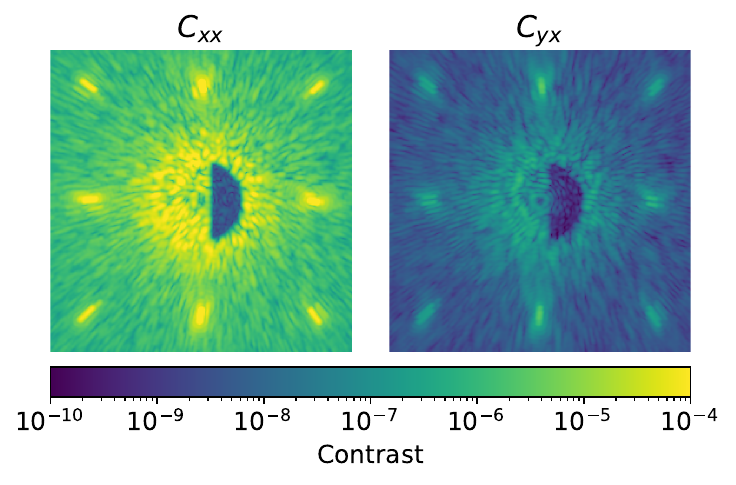}
    \caption{Simulated post-EFC broadband (10\% BW) contrast contribution from the Jones primary (Cxx) and cross-term (Cyx) electric field components for SCoOB}
    \label{fig:jones_contrast}
\end{figure}

Vector diffraction capability is added in a fork of \texttt{poppy}\cite{poppy_vector}, to be merged into the main branch in the upcoming months. Parametrized model generation and additional \texttt{poppy.OpticalElement} classes to enable the above analysis are included in the \texttt{poppy\_utils}\cite{poppy_utils} package. An example model output is shown in Figure \ref{fig:jones_contrast}.

\section{CONCLUSIONS}

We have developed analytic and numerical models to predict contrast limits set by lateral chromatic aberration in high contrast imaging systems, summarized in Equations \ref{eqn:c_wedge} and \ref{eqn:c_ppp}. With multi-band iEFC, we find that SCoOB faces a  contrast limit of \({\sim}4\times10^{-8}\) in a 10\% BW and \({\sim}3\times10^{-8}\) in a 5\% bandwidth in the presence of 30 arcminute wedged substrates sandwiching the circular polarizers. These wedge angles are significantly larger than required to move ghosts outside the $10 \lambda/D$ OWA, and future iterations will reduce these wedge angles to the 5 arcminute level to reduce the chromaticity in the system.

Future experiments will focus on tracking down the discrepancy between the modeled and actual broadband contrast floor. Work to reduce sources of stray light in the testbed is ongoing, and newly-acquired precision field stops will be integrated in the system to aid in this effort.

\appendix    

\acknowledgments 
Portions of this research were supported by funding from the Technology Research Initiative Fund (TRIF) of the Arizona Board of Regents and from Schmidt Sciences. J.N.A was supported by NASA through the NASA Hubble Fellowship grant \#HST-HF2-51547.001-A awarded by the Space Telescope Science Institute, which is operated by the Association of Universities for Research in Astronomy.

\bibliography{report} 
\bibliographystyle{spiebib} 

\end{document}